\documentclass[a4paper,11pt]{article}
\usepackage[usenames]{color}

\usepackage{amsfonts,amssymb}
\usepackage{theorem}
\usepackage[dvips]{graphicx}

\def\G{\Gamma}

\begin{document}

\rightline{IFUM-922-FT}

\vskip 0.3 truecm
\Large
\bf
\centerline{The $SU(2) \otimes U(1)$ Electroweak Model based 
}
\centerline{on the Nonlinearly Realized Gauge Group}
\normalsize \rm

\large
\rm
\vskip 0.3 truecm
\centerline{D.~Bettinelli$^{b,}$\footnote{e-mail: 
{\tt daniele.bettinelli@mi.infn.it}}, 
R.~Ferrari$^{a,b,}$\footnote{e-mail: {\tt ruggero.ferrari@mi.infn.it}}, 
A.~Quadri$^{b,}$\footnote{e-mail: {\tt andrea.quadri@mi.infn.it}}}

\normalsize
\medskip
\begin{center}
$^a$
Center for Theoretical Physics\\
Laboratory for Nuclear Science\\
and Department of Physics\\
Massachusetts Institute of Technology\\
Cambridge, Massachusetts 02139
and\\
$^b$ 
Dip. di Fisica, Universit\`a degli Studi di Milano\\
and INFN, Sez. di Milano\\
via Celoria 16, I-20133 Milano, Italy
\end{center}

\vskip 0.4  truecm
\normalsize
\centerline{{\bf Abstract}}
\rm
\begin{quotation}
The electroweak model is formulated on the nonlinearly 
 realized gauge group 
$SU(2) \otimes U(1)$. This implies that 
in perturbation theory no Higgs field is present.
The paper provides the effective action at the tree
level, the Slavnov Taylor identity (necessary for the
proof of physical unitarity), the local functional equation
(used for the control of the amplitudes involving
the Goldstone bosons) and the subtraction procedure
(nonstandard, since the theory is not power-counting
renormalizable). Particular attention is devoted to the
number of independent parameters relevant for the vector
mesons; in fact there is the possibility of introducing
two mass parameters. With this choice  the relation between
the ratio of the intermediate vector meson masses and the 
Weinberg angle depends on an extra free parameter.
\par
We briefly outline a method for dealing with $\gamma_5$
in dimensional regularization.
The model is formulated in the Landau gauge for sake of
simplicity and conciseness.
The QED Ward identity has a simple and intriguing form.

\end{quotation}

\newpage

\section{Introduction}
\label{sec:intr}

Within the framework of the nonlinearly realized gauge theories
of massive vector mesons, the extension of Yang-Mills theory 
from $SU(2)$ \cite{Bettinelli:2007tq}
to the $SU(2)\otimes U(1)$
group of the  electroweak model \cite{ew} is far from being
straightforward: the direction of the Spontaneous Symmetry
Breaking (SSB) and the dependence of the tree-level
action from this direction are non-trivial questions. 
The pure $SU(2)$ Yang-Mills theory
\cite{Yang:1954ek}, when a mass term is introduced by 
the nonlinearly realization of the gauge group, requires
the use of transformations of {\sl local} $SU(2)_L$ left and
of {\sl global} $SU(2)_R$ right{; the local functional equation
associated to the $SU(2)_L$ invariance together with the Weak
Power Counting (WPC) allows to overcome the problem of nonrenormalizability
and of the anomalous interaction terms, while the
$SU(2)_R$ selects a single symmetric mass term 
\cite{Bettinelli:2007tq}.}
The introduction of a
$U(1)$ associated to the hypercharge destroys the {\sl global} 
$SU(2)_R$ right symmetry. {Thus  two mass
invariants  can be introduced}:
one for the neutral and one for the charged vector meson.
Consequently the ratio of the vector meson masses is not fixed anymore 
by the Weinberg angle.

In this paper we start from the 
unique tree-level vertex functional of the theory
 based on the nonlinear realization of the 
$SU(2) \otimes U(1)$ gauge symmetry
in the presence of the fermionic matter content
of the Standard Model (with massless neutrinos) and
compatible with the WPC.
This provides the Feynman rules in terms of the
tree level parameters and the overall mass scale $\Lambda$ 
{for the radiative corrections}. 
{A whole set of external sources are used
in order to introduce a closed set of local operators necessary
for the Becchi-Rouet-Stora-Tyutin (BRST) transformations \cite{brst},
the local $SU(2)_L$ transformations and the Landau gauge fixing.
The functional equations derived from the invariance of the
path integral measure are the tools for the construction
of the theory: the Slavnov-Taylor identities (STI) \cite{ST}
in order to prove physical unitarity \cite{physunit}, \cite{Ferrari:2004pd},
the Local Functional Equation (LFE) 
\cite{Ferrari:2005ii}, \cite{Bettinelli:2007tq} for the symmetric
subtraction procedure and control of the tree-level couplings
by the WPC \cite{Ferrari:2005va}, 
\cite{Bettinelli:2007tq} 
and the Landau gauge equation.
}
\par
It is understood that we approach the quantization by a series
expansion in $\hbar$. Thus no elementary field for the Higgs
Boson \cite{ssb} is present in the theory, since the representation is
nonlinear. The scenario is then very interesting. For instance
an Higgs boson could emerge as a non-perturbative mechanism,
but then its physical parameters are not constrained by the
radiative corrections of the low energy electroweak processes.
Otherwise, our energy scale for the radiative corrections  
$\Lambda$ is a manifestation of {some other}
high energy physics.
\par
The intention of the present note is to provide the theoretical
basis and the technical tools for explicit calculations
in the electroweak model based on the  nonlinearly realized
gauge group. 
Our subtraction procedure is based on minimal subtraction
on specifically normalized amplitudes. Thus the presence of 
$\gamma_5$ poses serious problems to the whole strategy. Our
approach is pragmatic. {A new  $\gamma_{_D}$ is introduced,}
which anticommutes with every $\gamma_\mu$, and at the same time
no statement is made about the analytic properties of
the trace involving $\gamma_{_D}$. Since the theory is not anomalous
such traces never meet poles in $D-4$ and therefore we
can impose that their limit for $D=4$ is continuous. 
\par
In this paper the Landau gauge is used for sake of simplicity
and conciseness. Other covariant gauges are possible as
discussed in ref.  \cite{Bettinelli:2007eu}
for pure SU(2) massive Yang-Mills.
\par
We postpone to a future publication the  {details in the derivation}
of the LFE \cite{Ferrari:2005ii},\cite{Bettinelli:2007zn}. 
The use of LFE to establish the hierarchy among the Green
functions and the control of the divergences in the limit $D=4$
are described in Refs. \cite{Ferrari:2005ii} and \cite{Ferrari:2005fc}.
The geometrical aspects and the solutions of the linearized LFE
can be found in \cite{Bettinelli:2007kc}.
The classical action is proposed in this paper, by using the criterion
of WPC introduced 
for the nonlinear sigma model \cite{Ferrari:2005va} and 
used  in massive SU(2) Yang-Mills model \cite{Bettinelli:2007tq}.
The use of the Slavnov-Taylor identity in the case of Landau
gauge in order to guarantee physical unitarity is discussed in ref.
\cite{Ferrari:2004pd}.
\par
Some very important issues are not discussed here. In particular
the connection of the present approach with previous attempts 
to remove the Higgs contribution in the large mass limit
as in \cite{Dittmaier:1995cr}. A similar approach for the
nonlinear sigma model turns out to be exceedingly complex 
\cite{Bettinelli:2006ps}. The issue of unitarity at large
energy \cite{Froissart:1961ux} when the Higgs field
is removed (as in ref. \cite{Lee:1977eg})  is not discussed here. 
\par
In the present paper we are going to propose a subtraction procedure
of the divergences which is unique; i.e. there are no extra free
parameters originated by the subtraction procedure of the divergences,
beside the mass scale of the radiative corrections 
$\Lambda$.
Since the counterterms associated to finite renormalization
cannot be reinserted back in the tree-level vertex functional
without  violating either the symmetries or the WPC, we 
cannot perform on-shell finite renormalizations,
since this implies finite counterterms at every order in the
perturbative expansion. Moreover it turns out that in this scheme
the symmetric formalism is the most practical one at variance
with the formalism based on physical fields. 
In particular the
use of the fields in the symmetric basis and of the Landau gauge
yields a very simple and intruiging form for the Ward identity
generated by the electric charge.

\normalsize
\medskip
%
%
\section{Preliminaries}
\label{sec:prel}

We start from the classical action without gauge fixing and external
sources for the composite operators. This allows a simpler
discussion about the number of parameters and the constraint of
WPC. In the next sections we shall introduce the
gauge fixing and the external sources. Some details will be written 
in a formalism that generalizes the conventional notation.
\begin{eqnarray}&&
\Gamma^{(0)}=
\Lambda^{(D-4)} \int d^Dx\,\Biggl( \,2 \,Tr\, \biggl\{
- \frac{1}{4}  G_{\mu\nu} G^{\mu\nu} - \frac{1}{4}  F_{\mu\nu}
F^{\mu\nu}
\Biggr\}
\nonumber\\&&
+M^2  \,Tr\, \biggl\{\bigl (gA_{\mu}
- \frac{g'}{2} \Omega\tau_3 B_\mu \Omega^\dagger
- F_{\mu}\bigr)^2\biggr\}
\nonumber\\&& 
+M^2\frac{ \kappa }{2}\Bigl( Tr\bigl\{(g \Omega^\dagger A_\mu \Omega 
- g' B_\mu \frac{\tau_3}{2}
+ i \Omega^\dagger \partial_\mu \Omega) \tau_3\bigr\}\Bigr)^2
\nonumber\\&&
+\sum_L\biggl[
\bar L \bigr(i\not\!\partial +g\not\!\!A 
+\frac{g'}{2}Y_L\not\!B\bigl)L
+\sum_R\bar R \bigr(i\not\!\partial 
+ \frac{g'}{2} (Y_L+\tau_3) \not\!B\bigl)R
\biggr]
\nonumber \\&&
+\sum_j\biggl[
m_{l_j}~\bar R^l_j\frac{1-\tau_3}{2}\Omega^\dagger L^l_j
-
m_{q^u_j}~\bar R^q_j \frac{1+\tau_3}{2}\Omega^\dagger L^q_j
\nonumber\\&&
+
m_{q^d_k} V^\dagger_{kj} ~\bar R^q_k
\frac{1-\tau_3}{2}\Omega^\dagger 
 L^q_j +h.c.
\biggr]
\Biggr)
\label{pre.1}
\end{eqnarray}
where $L$ and $R$ are doublets such that
\begin{eqnarray}
\gamma_{_D} L = - L \quad
\gamma_{_D} R =  R,
\label{pre.2}
\end{eqnarray}
being $\gamma_{_D}$ a gamma matrix that anticommutes with every other
$\gamma^\mu$.
$Y_L$ is the hypercharge of the left-fields.
We use also the   $2\times 2$ matrix notation for the fields  
$A_\mu,\Omega, F_\mu$ 
\begin{eqnarray}&&
A_{\mu}= A_{a\mu} \frac{\tau_a}{2}\qquad 
\Omega = \frac{1}{v}(\phi_0+i\tau_a \phi_a),\quad\Omega \in SU(2)
\nonumber \\&&
G_{\mu\nu}=\partial_\mu A_\nu- \partial_\nu A_\mu -ig[A_\mu,A_\nu]
\nonumber \\&&
F_{\mu} =  i\Omega\partial_\mu \Omega^\dagger 
= F_{a\mu} \frac{\tau_a}{2}
\nonumber \\&&
F_{\mu\nu}=\partial_\mu B_\nu- \partial_\nu B_\mu.
\label{pre.3}
\end{eqnarray}
\subsection{Fermions}
The quark fields $(q^u_j,j=1,2,3)=(u,c,t)$ and $( q^d_j,j=1,2,3)=(d,s,b)$ 
are taken to be the mass eigenstates in the tree level lagrangian. Similar
notation is used for the leptons $(l^u_j,j=1,2,3)=(\nu_e,\nu_\mu,\nu_\tau)$ and 
$( l^d_j,j=1,2,3)=(e,\mu,\tau)$.
$L$ is an element of the set of the left fields of the three families
\begin{eqnarray}
L  \in\Biggl\{ \left(
\begin{array}{r} l^u_{Lj}\\
l^d_{Lj}
\end{array} \right), \left(
\begin{array}{r}q^u_{Lj}\\
V_{jk}q^d_{Lk}
\end{array} \right), \quad j,k=1,2,3\Biggr\},
\label{pre.4}
\end{eqnarray} 
where $V_{jk}$ is the CKM matrix;
the right components
can also be written formally as doublets
\begin{eqnarray}
R  \in\Biggl\{ \left(
\begin{array}{r}l^u_{Rj} \\
l^d_{Rj}
\end{array} \right), \left(
\begin{array}{r} q^u_{Rj}\\
q^d_{Rj}
\end{array} \right), \quad j,k=1,2,3\Biggr\}
\label{pre.5}
\end{eqnarray} 
(color indices are not exhibited).
The single left doublets are denoted by $L^l_j$,
$j=1,2,3$ for the leptons, $L^q_j$, $j=1,2,3$ for the 
quarks.
\subsection{$SU(2)$ left - and $U(1)$ right-local transformations}
By making use of the path integral we shall derive
some identities which stem both from the invariance
of the integration measure over the fields and
from the transformation properties of the action.
If the action is not invariant, then one has to add
new source terms coupled to the new generated 
operators. Thus we study the invariance properties of the functional 
in eq. (\ref{pre.1})  under the $SU(2)$ local
transformations, where $\Omega$ is transformed on the left (thus
we use the notation $SU(2)_L$)
\begin{eqnarray}
\begin{array}{ll}
\Omega' = U\Omega &\quad B'_\mu = B_\mu\\
A'_{\mu}= UA_{\mu}U^\dagger + \frac{i}{g} U\partial_\mu U^\dagger
&\quad L' = UL \\
F'_{\mu}= UF_{\mu}U^\dagger + i U\partial_\mu U^\dagger&\quad
R' = R
\end{array}.
\label{pre.6}
\end{eqnarray}
Under the $\exp (i\frac{\alpha}{2} Y)\in U(1)$  local transformations
$\Omega$ is transformed on the right (thus
we use the notation $U(1)_R$)
\begin{eqnarray}
\begin{array}{ll}
e^{-i\frac{\alpha}{2} Y}\Omega e^{i\frac{\alpha}{2} Y}
 = \Omega V^\dagger &\quad
e^{-i\frac{\alpha}{2} Y} F_{\mu} e^{i\frac{\alpha}{2} Y}
   =   F_{\mu} + i \Omega V^\dagger\partial_\mu V \Omega^\dagger\\
e^{-i\frac{\alpha}{2} Y} A_{\mu}  e^{i\frac{\alpha}{2} Y}
 = A_{\mu} &\quad
e^{-i\frac{\alpha}{2} Y} B_\mu e^{i\frac{\alpha}{2} Y}
 = B_\mu{+} \frac{1}{g'} \partial_\mu \alpha\\
e^{-i\frac{\alpha}{2} Y}L  e^{i\frac{\alpha}{2} Y}
   = \exp(i\frac{\alpha}{2} Y_L) L
 &\quad
e^{-i\frac{\alpha}{2} Y}R  e^{i\frac{\alpha}{2} Y}
   = \exp(i\frac{\alpha}{2} (Y_L+\tau_3)) R,
\end{array}
\label{pre.7}
\end{eqnarray}
where $Y_L$ is the hypercharge of the $L$ and
\begin{eqnarray}
V(\alpha)=\exp (i\frac{\alpha}{2}\tau_3).
\label{pre.8}
\end{eqnarray}
The nonlinearity of the representation comes from the
constraint
\begin{eqnarray}
\Omega\Omega^\dagger =1 \Longrightarrow \phi_0^2+\vec\phi^2=v^2.
\label{pre.8.1}
\end{eqnarray}
\par
The electric charge is defined as usual
\begin{eqnarray}
Q= I_3+\frac{1}{2}Y
\label{pre.8.2}
\end{eqnarray}
where $I_3$ is a generator of $SU(2)_L$. Since the symmetry generated by $Q$
is not spontaneously broken, then the component $\Omega_0$ which 
acquires a non zero  vacuum expectation value
 must obey the condition
\begin{eqnarray}
\langle 0|[Q,\Omega]|0\rangle=0 \Longrightarrow \frac{1}{2}\tau_3\Omega_0
-\frac{1}{2}\Omega_0 \tau_3
=0.
\label{pre.8.3}
\end{eqnarray}

Out of the manifold that solves (\ref{pre.8.3}) we choose, at the tree
level, the direction
\begin{eqnarray}
\phi_0=\frac{1}{v}\sqrt{v^2-\vec\phi^2}.
\label{pre.8.4}
\end{eqnarray}
The choice might not be stable under radiative corrections. However
in our present approach (no Higgs and 
Landau gauge) the absence of tadpole graphs indicates that
the vacuum expectation value of $\phi_0$ receives no radiative corrections.
%
\subsection{Two mass invariants and the WPC}
The two expressions in eq. (\ref{pre.1}) multiplied
by $M^2$ are invariant under $SU_L(2)\otimes U(1)_R$
transformations given by eqs. (\ref{pre.6}) and (\ref{pre.7}) 
\cite{massinv};
in fact the bleached field
\begin{eqnarray}
w_\mu=\Omega^\dagger gA_{\mu}\Omega
-g' \frac{\tau_3}{2}  B_\mu +i\Omega^\dagger\partial_\mu \Omega
\label{pre.9}
\end{eqnarray}
is an $SU(2)_L$-invariant and transforms according to
\begin{eqnarray}
w_\mu'= Vw_\mu V^\dagger.
\label{pre.10}
\end{eqnarray}
For each $\mu$ $w_\mu$ has four components. However, since
\begin{eqnarray}
w_\mu^\dagger =  w_\mu, \qquad 
Tr \{w_\mu\}=0,
\label{pre.11}
\end{eqnarray}
one gets
\begin{eqnarray}
(w_\mu)_{11}=-(w_\mu)_{22} \qquad (w_\mu)_{12}^*=(w_\mu)_{21}.
\label{pre.11.1}
\end{eqnarray}
Therefore the only two independent mass invariants are $(w_\mu)_{11}^2$
and $(w^\mu)_{21} (w_\mu)_{12}$.
\par
Let us introduce the notation
\begin{eqnarray}
\Phi \equiv \left(
\begin{array}{l}
i\phi_1+\phi_2\\ \phi_0-i\phi_3
\end{array}
\right), \quad
\Phi^c\equiv i\tau_2
\Phi^* \equiv \left(
\begin{array}{l}
\phi_0+i\phi_3\\i\phi_1-\phi_2
\end{array}
\right);
\label{pre.12}
\end{eqnarray}
hence 
\begin{eqnarray}
\Omega_{\alpha\beta}=\frac{1}{v}\Phi^c_\alpha\Phi_\beta.
\label{pre.13}
\end{eqnarray}
From eqs. (\ref{pre.6}) and  (\ref{pre.7}) one derives the
transformation properties under $SU(2)_L\otimes U(1)_R$
\begin{eqnarray}
\Phi'= U\Phi ~e^{-i\frac{\alpha}{2}}.
\label{pre.14}
\end{eqnarray}
In eq. (\ref{pre.1}) the first mass invariant can be
written in the form \cite{st}
\begin{eqnarray}
2\frac{M^2}{v^2}\biggl|\biggl(gA_\mu   - \frac{g'}{2}  B_\mu
+i\partial_\mu\biggr) \Phi
\biggr|^2
\label{pre.15}
\end{eqnarray}
while the second \cite{massinv}
\begin{eqnarray}
2\kappa \frac{M^2}{v^4}\biggl|\Phi^\dagger\biggl(gA_\mu   - \frac{g'}{2}  B_\mu
+i\partial_\mu\biggr) \Phi
\biggr|^2.
\label{pre.16}
\end{eqnarray}
In the mass eigenstate basis given by
\begin{eqnarray}
W_\mu^\pm = \frac{1}{\sqrt{2}} (A_{1\mu} \mp i A_{2\mu}) 
\label{wpm}
\end{eqnarray}
and
\begin{eqnarray}&& 
Z_\mu = \frac{1}{\sqrt{g^2+g^{'2}}}(g A_{3\mu}-g' B_\mu)
\nonumber\\&&
A_\mu = \frac{1}{\sqrt{g^2+g^{'2}}}(g' A_{3\mu}+g B_\mu),
\label{charge.2.mass}
\end{eqnarray}
one finds 
\begin{eqnarray}
M_W^2 = g^2 M^2 \, , ~~~~ M_Z^2 = (1+\kappa) 
\frac{g^2 M^2}{c^2} \, ,
\label{mass.inv.1}
\end{eqnarray}
while the photon $A_\mu$ is massless.

In the eq.(\ref{mass.inv.1}) $c$ is the cosine of the Weinberg
angle $\theta_W$. The latter 
 is defined as usual according to 
\begin{eqnarray}
\tan \theta_W = \frac{g'}{g} \, .
\label{wein.angle}
\end{eqnarray}
The ratio of $M_W$ and $M_Z$ 
in eq.(\ref{mass.inv.1})
is a function of the
parameter $\kappa$. 
This is a peculiar feature of the nonlinearly realized
electroweak model.

The invariance properties of the expressions in eqs. 
(\ref{pre.15}) and (\ref{pre.16}) are valid
independently from the constraint (\ref{pre.8.1}). Thus the technique
of bleaching for the construction of the $SU(2)\otimes U(1)$
invariants
can be freely used also for the case of linear representation 
(Higgs mechanism).
However, in renormalizable theories the term in eq. (\ref{pre.16}) is
excluded being of dimension 6.
The fermion sector can be considered in the same way by using the
bleaching
\begin{eqnarray}
{\tilde l^d_{_{Lj}}}\equiv \frac{1}{v}\Phi^{\dagger} l_{_{Lj}},\quad 
\tilde q^u_{_{Lj}}\equiv \frac{1}{v}\Phi^{c\dagger}  q_{_{Lj}},\quad
\tilde q^d_{_{Lj}}\equiv \frac{1}{v}\Phi^{\dagger} q_{_{Lj}}.
\label{pre.17}
\end{eqnarray}
\par
In our case (nonlinear representation of the gauge group)
the generic graph with no external Goldstone boson legs
(ancestor amplitudes) has a degree of superficial 
divergence bounded by
\begin{eqnarray}
d(G) \leq (D-2)n_L +2 -N_B -N_F
\label{pre.18}
\end{eqnarray}
where $n_L$ is the number of loops and $N_B,N_F$ the number
of external gauge- and fermion-fields. Thus in the limit $D=4$ 
the number of divergent one-particle-irreducible (1-PI) amplitudes 
is finite at fixed number of loops; while, if we consider
also external Goldstone boson legs (descendant amplitudes),
already at one loop the number of divergent  (1-PI) amplitudes
is infinite. It is interesting that the Fermions enter in eq.
(\ref{pre.18}) with dimension $1$ instead of the canonical $\frac{3}{2}$.
\par
By imposing this constraint (WPC) one obtains that
the number of independent and invariant terms in the action
are those exhibited in eq. (\ref{pre.1}). Moreover it
can be shown that our subtraction procedure, described in the sequel,
does not destroy the bound given in eq. (\ref{pre.18}).
Thus the WPC becomes a very important and efficient tool
in the construction of the classical action of the model.
In particular the action is protected against anomalous couplings,
which are present if one relays only on symmetry requirements
\cite{Peccei:1989kr}.

\section{Quantization, gauge fixing, ST, LFE, 
Landau gauge equation}
\label{sec:quan}
The classical action in eq. (\ref{pre.1}) is the
starting point of a complex strategy that takes into account
the field quantization (which we perform with the
tool of a gauge fixing). Due to the presence
of unphysical modes, one has to introduce some
Faddeev-Popov ghosts by requiring BRST invariance.
The STI then translate on
the Feynman amplitudes the BRST invariance and
provide  the necessary relations
that guarantees the cancellation of the contributions
of the unphysical modes for physical amplitudes.
A further local functional equation (LFE) is necessary
to account for the fact that the Goldstone modes
enter as gauge modes in the model. The LFE allows
to trace the correct subtraction procedure for
the divergences and moreover it guarantees full
hierarchy, i.e. all amplitudes involving the
Goldstone boson (descendant amplitudes) are derived
from the ancestor amplitudes (i.e. with no Goldstone
bosons). The Landau gauge equation follows
from the gauge fixing term and it is equivalent
to the anti-ghost equation by making use of the STI.
\par
According to this procedure we start from the
classical action, add the gauge fixing terms and
the Faddeev-Popov ghosts in order to implement
the BRST transformations. We have to introduce all
the source terms necessary for the renormalization
of the new necessary composite operators. When we consider
the $SU(2)_L\otimes U(1)_R$ transformations, again new
composite operators emerge. We provide a set
of source terms that is closed under the combined
set of transformations. The complete analysis
is left to a future work. Here we give the final
results.
\par
The tree level effective action  describing the gauge fixing 
(Landau gauge) and the composite operator source terms is
(in the notation we indicate with a $^*$ the sources necessary
for the formulation of the STI: the
anti-fields.)
\begin{eqnarray}&&
\Gamma^{(0)}_{\rm GF}
\nonumber\\&&
=
\Lambda^{(D-4)}\int d^Dx \Biggl( b_0 \partial_\mu B^\mu 
-\bar c_0 \Box c_0 +2 Tr~\Bigl\{  b\partial_\mu A^\mu 
-\bar c\partial^\mu D[A]_\mu c
\nonumber\\&&
+ V^\mu ~\biggl ( D[A]_\mu b 
-ig \bar cD[A]_\mu c - ig (D[A]_\mu c) \bar c
\biggr)
+\Theta^\mu~D[A]_\mu\bar c\Bigr\} +K_0\phi_0 
\nonumber\\&& 
+ A^*_{a\mu} \mathfrak{s}A^\mu_a + 
\phi_0^* \mathfrak{s} \phi_0 + 
\phi_a^* \mathfrak{s} \phi_a + c_a^* \mathfrak{s} c_a
+ \sum_L\Big ( L^* \mathfrak{s}L + 
\bar L^* \mathfrak{s}\bar{ L}\Bigr)
\Biggr) \, ,
\label{pre.19}
\end{eqnarray}
where the Lagrange multipliers and the ghosts of $SU(2)_L$ are
in matrix notation
\begin{eqnarray}&&
b= b_a\frac{1}{2}\tau_a, \qquad  c= c_a\frac{1}{2}\tau_a
, \qquad \bar  c=\bar  c_a\frac{1}{2}\tau_a
\nonumber\\&&
D[A]_\mu ~ c = \partial_\mu~ c- ig[A_\mu, c]
\nonumber\\&&
=     \frac{1}{2} \tau_a    
( \partial_\mu \delta_{ab}- g\epsilon_{abc} A_{c\mu})c_b.
\label{prel.7.nn}
\end{eqnarray}
The full effective action at the tree level (eqs. (\ref{pre.1}) 
and (\ref{pre.19})) is invariant
under the BRST transformations
(not counting the source terms)
\begin{equation}
\begin{array}{llll}
\mathfrak{s} A_\mu = D[A]_\mu ~ c
&
\mathfrak{s} \Omega = ig~ c ~ \Omega~~~
&
\mathfrak{s} \bar c = b 
&
\mathfrak{s} \bar c_0 = 0 
\\
\mathfrak{s} c = i g~ c ~ c
&
\mathfrak{s} B_\mu = 0
&
\mathfrak{s} b = 0
&
\mathfrak{s} b_0 = 0
\\
\mathfrak{s}L = igcL
&
\mathfrak{s}R =0
&
\mathfrak{s}c_0=0
&
\end{array} 
\label{pre.20}
\end{equation}
\begin{equation}
\begin{array}{llll}
\mathfrak{s}_1 A_\mu = 0
&
\mathfrak{s}_1 \Omega =- \frac{i}{2} g' \Omega c_0 \tau_3
&
\mathfrak{s}_1 \bar c=0
&
\mathfrak{s}_1 \bar c_0 = b_0
\\

\mathfrak{s}_1 c=0
&
\mathfrak{s}_1 B_\mu = \partial_\mu  c_0
&
\mathfrak{s}_1 b=0
&
\mathfrak{s}_1 b_0 = 0
\\
\mathfrak{s}_1 L= \frac{i}{2} g'c_0Y_LL
&
\mathfrak{s}_1 R = \frac{i}{2} g'c_0(Y_L+\tau_3) R
&
\mathfrak{s}_1 c_0 = 0.
&
\end{array} 
\label{pre.21}
\end{equation}
By construction 
\begin{eqnarray}
\{\mathfrak{s},\mathfrak{s}_1\}=0.
\label{pre.22}
\end{eqnarray}
No sources for the $\mathfrak{s}_1$ transforms of fields are used, since
$c_0$ is a free field.
\par
From the BRST transformation (\ref{pre.20}) we get the STI
\footnote{The notation is as follows: $\G_\psi$ stands
for $\delta \G / \delta \psi$, while $W_\psi$ 
for $\delta W / \delta J_\psi$, with $\psi$ any of the quantized
fields of the model and $J_\psi$ its source. The connected
generating functional $W[J]$ is related to the vertex functional
$\G[\psi]$ by $W=\G+ \int d^Dx \, J \psi$ \, .}
\begin{eqnarray}
&& \!\!\!\!\!\!\!\!\!\!
{\cal S}\G \equiv \int d^Dx \,\Biggl[ \Lambda^{-(D-4)}\Big (
 \G_{ A^*_{a\mu}}  \G_{ A_a^\mu}
+
 \G_{ \phi_a^*}  \G_{ \phi_a}
+ 
 \G_{ c_a^*} \G_{ c_a}
 \nonumber \\
&&
+ \G_{ L^*}  \G_{ L}
+ \G_{ \bar L^*} 
 \G_{ \bar L}\Big )
+ b_a  \G_{ \bar c_a}
 + \Theta_{a\mu}  \G_{ V_{a\mu}}
      - K_0  \G_{ \phi_0^*} 
\Biggr] = 0. 
\label{brst.13}
\end{eqnarray}
The classical linearized form of the
operator is
\begin{eqnarray}
&& \!\!\!\!\!\!\!\!\!\!
{\cal S}_0\G \equiv \int d^Dx \,\Biggl[ \Lambda^{-(D-4)}\Big (
\G^{(0)}_{ A_a^\mu} 
\frac{\delta }{\delta A^*_{a\mu}}
+
\G^{(0)}_{ A^*_{a\mu}} \frac{\delta}{\delta A_a^\mu}
+
\G^{(0)}_{ \phi_a^*} \frac{\delta }{\delta \phi_a}
+
 \G^{(0)}_{ \phi_a}\frac{\delta }{\delta \phi_a^*}
 \nonumber \\
&&
+ 
\G^{(0)}_{ c_a^*}\frac{\delta }{\delta c_a}
+ 
\G^{(0)}_{ c_a}\frac{\delta }{\delta c_a^*}
+\G^{(0)}_{ L^*} \frac{\delta }{\delta L}
+ \G^{(0)}_{ L}\frac{\delta }{\delta L^*}
 \nonumber \\
&&
+\G^{(0)}_{ \bar L^*} 
\frac{\delta }{\delta \bar L}
+\G^{(0)}_{ \bar L}
\frac{\delta }{\delta \bar L^*} \Big )
+ b_a \frac{\delta }{\delta \bar c_a}
 + \Theta_{a\mu} \frac{\delta }{\delta V_{a\mu}}
      - K_0 \frac{\delta }{\delta \phi_0^*} 
\Biggr]\G .
\label{brst.14}
\end{eqnarray}
In both eqs. (\ref{brst.13}) and (\ref{brst.14}) the sum over
$L$ and $\bar L^*$ is understood over the components
explicitly shown in eq. (\ref{pre.4}).
From the transformations in eq. (\ref{pre.21}) we get the 
relation
\begin{eqnarray}
&&
{\frac{2}{g'}}\Box b_0
- {\frac{2}{g'}}\partial^\mu  \frac{\delta\G }{\delta  B^\mu}
+{\Lambda^{(D-4)}}\phi_3 K_0 
+ \phi_2 \frac{\delta\G }{\delta \phi_1}
-\phi_1\frac{\delta\G }{\delta \phi_2}
-{\frac{1}{\Lambda^{(D-4)}}}
\frac{\delta\G }{\delta K_0}\frac{\delta\G }{\delta \phi_3}
\nonumber\\&&
-\phi_3^*\frac{\delta\G }{\delta \phi_0^*}
+ \phi_2^* \frac{\delta\G }{\delta \phi_1^*}
-\phi_1^* \frac{\delta\G }{\delta \phi_2^*}
+\phi_0^*\frac{\delta\G }{\delta \phi_3^*} 
\nonumber\\&&
+iY_LL\frac{\delta\G }{\delta L}
-iY_L\bar L\frac{\delta\G }{\delta \bar L}
+i(Y_L+\tau_3)R\frac{\delta\G }{\delta R}
-i\bar R(Y_L+\tau_3)\frac{\delta\G }{\delta \bar R}
\nonumber\\&&
-iY_LL^*\frac{\delta\G }{\delta L^*}
+iY_L\bar L^*\frac{\delta\G }{\delta \bar L^*}
=0.
\nonumber\\&&
\label{st.3}
\end{eqnarray}
%
%
%
The same relation is obtained by using the $U(1)_R$ transformations
of eq. (\ref{pre.7}), complemented with the following extension
to the new variables 
\begin{equation}
\begin{array}{llll}
V_\mu' = V_\mu 
&
\Omega^{*'} = {V\Omega^* }
&
L^{*'} = \exp(-i\frac{\alpha}{2} Y_L) L^*
&
K_0' = K_0
\\
\Theta_\mu'=\Theta_\mu
&
b'=b
&
\bar L^{*'} = \exp(i\frac{\alpha}{2} Y_L) \bar L^*
&
 b_0' = b_0
\\
c'=c
&
\bar c' = \bar c
&
c^{*'}=c^*
&
\\ 
c_0' = c_0
&
\bar c_0' = \bar c_0
&
A_\mu^{*'}= A_\mu^{*}.
&
\end{array} 
\label{st.3.1}
\end{equation}
The Landau gauge equation is
\begin{eqnarray}
\G_{ b_a} =  \Lambda^{(D-4)} \Big ( 
 D^\mu[V](A_\mu - V_\mu)\Big )_a 
\label{b.eq}
\end{eqnarray}
which implies the ghost equation
\begin{eqnarray}
 \G_{ \bar c_a} =  
 \Big ( 
-D_\mu[V]  \G_{ A_{\mu}^*} 
+  \Lambda^{(D-4)}  D_\mu[A] \Theta^\mu
\Big )_a \, ,
\label{gh.eq}
\end{eqnarray}
by using the STI (\ref{brst.13}).
\par
Now we explore the LFE that follows
from the invariance of the path integral measure 
over the transformations (\ref{pre.6}). In doing that
we have first to extend the transformations to the newly
introduced fields and sources. This is straightforward by
following the criterion that all
the transformations should close on a finite number of
composite operators. Thus eq. (\ref{pre.6}) is complemented
by
\begin{equation}
\begin{array}{llll}
V'_{\mu}= UV_{\mu}U^\dagger + \frac{i}{g} U\partial_\mu U^\dagger
&
\Omega^{*'} = {\Omega^*U^\dagger }
&
L^{*'} =  L^*U^\dagger
&
K_0' = K_0
\\
\Theta_\mu'=U\Theta_\mu U^\dagger
&
b'=UbU^\dagger
&
\bar L^{*'} = U \bar L^*
&
 b_0' = b_0
\\
c'=Uc U^\dagger
&
\bar c' = U \bar c U^\dagger
&
c^{*'}=c^*
&
\\ 
c_0' = c_0
&
\bar c_0' = \bar c_0
&
A_\mu^{*'}= UA_\mu^{*} U^\dagger.
&
\end{array} 
\label{pre.6.1}
\end{equation}
Thus the resulting identity associated to the $SU(2)_L$
local transformations is ($x$-dependence is not shown)
\begin{eqnarray}
&&({\cal W}\G)_{a} \equiv 
-{\frac{1}{g}}\partial_\mu \G_{ V_{a \mu}} 
+ \epsilon_{abc} V_{c\mu} \G_{ V_{b\mu}}
-{\frac{1}{g}}\partial_\mu \G_{ A_{a \mu}}
\nonumber \\&&  
+ \epsilon_{abc} A_{c\mu} \G_{ A_{b\mu}}
+ \epsilon_{abc} b_c \G_{ b_b}
+ \frac{{\Lambda^{(D-4)}}}{2} K_0\phi_a
+ \frac{1}{2\Lambda^{(D-4)}} \G_{ K_0} 
\G_{ \phi_a} 
\nonumber \\
&&  
+  
\frac{1}{2} \epsilon_{abc} \phi_c \G_{ \phi_b} 
      + \epsilon_{abc} \bar c_c \G_{ \bar c_b}
      + \epsilon_{abc} c_c \G_{  c_b}
 \nonumber \\&&
{
+\frac{i}{2}\tau_aL\G_{  L}
-\frac{i}{2}\bar L\tau_a\G_{ \bar  L}
-\frac{i}{2}L^*\tau_a\G_{  L^*}
+\frac{i}{2}\tau_a\bar L^*\G_{ \bar  L^*}
}
 \nonumber \\
&& + \epsilon_{abc} \Theta_{c\mu} \G_{ \Theta_{b\mu} }
      + \epsilon_{abc} A^*_{c\mu} \G_{ A^*_{b\mu}}
      + \epsilon_{abc} c^*_c \G_{  c^*_b} 
 - \frac{1}{2} \phi_0^* \G_{ \phi^*_a}
\nonumber \\
&& +  
\frac{1}{2} \epsilon_{abc} \phi^*_c \G_{ \phi^*_b} 
+ \frac{1}{2} \phi_a^* \G_{ \phi_0^*}
= 0 \, ,
\label{bkgwi}
\end{eqnarray}
where the nonlinearity of the realization of the $SU(2)_L$ 
gauge group is revealed by the presence of the bilinear term
{$ \G_{ K_0} \G_{ \phi_a} $}.
This is essential for the hierarchy; in fact eq. (\ref{bkgwi}) shows
that every amplitude with $\phi-$external  leg 
(descendant amplitudes) can be obtained from those without. 
When this property is 
supplemented with the WPC, it appears that the number of independent
counterterms necessary to make the theory is finite
at every order of the perturbative expansion in loops.
A quick inspection to eq. (\ref{bkgwi}) shows that one has to evaluate
a whole set of amplitudes with all possible external sources
in order to fix the descendant amplitudes. 
Moreover equation (\ref{pre.18}) have to be updated to the presence
of the external sources. A straightforward argument shows that
the superficial degree  of divergence of an ancestor
amplitude is bounded by
\begin{eqnarray}
&& d({\cal G}) \leq (D-2)n +2 - N_A - N_B - N_c - N_F 
- N_{\bar F}  -  N_V - N_{\phi_a^*} \nonumber \\
&& ~~~~ - 2 (N_\Theta + N_{A^*} + N_{\phi_0^*} + N_{L^*} + 
N_{{\bar L}^*} + N_{c^*} + N_{K_0} ) \, .
\label{pre.18p}
\end{eqnarray}
where $N_X$ is the number of external fields $X=A_\mu,B_\mu,b,b_0,L,R $
and sources $X=V_\mu,\Theta_\mu,A^*,L^*,\bar L^*,c^*,K_0$.
In passing it is worth
noticing that the STI are not sufficient to
fix all the descendant amplitudes. This feature is present also in 
pure $SU(2)$ massive Yang-Mills \cite{Bettinelli:2007tq}.
In conclusion, the LFE (\ref{bkgwi}) is
the right tool to describe at the quantum level the {\sl gauge}
character of the Goldstone boson fields $\vec\phi$ and how, through
the hierarchy and the bleaching technique, they can be managed.
\par
The classical linearized form of the operator in eq. (\ref{bkgwi}) is
\begin{eqnarray}
&&({\cal W}_0\G)_a \equiv \Biggl(
-{\frac{1}{g}}\partial_\mu \frac{\delta }{\delta V_{a \mu}} 
+ \epsilon_{abc} V_{c\mu} \frac{\delta }{\delta V_{b\mu}}
-{\frac{1}{g}}\partial_\mu \frac{\delta }{\delta A_{a \mu}}
\nonumber \\&&  
+ \epsilon_{abc} A_{c\mu} \frac{\delta }{\delta A_{b\mu}}
+ \epsilon_{abc} b_c \frac{\delta }{\delta b_b}
+ \frac{1}{2\Lambda^{(D-4)}}   \frac{\delta \G^{(0)}}{\delta K_0}
\frac{\delta }{\delta \phi_a}
\nonumber \\
&&  
+ \frac{1}{2\Lambda^{(D-4)}} 
\frac{\delta \G^{(0)}}{\delta \phi_a} 
\frac{\delta }{\delta K_0} 
+  
\frac{1}{2} \epsilon_{abc} \phi_c \frac{\delta }{\delta \phi_b} 
      + \epsilon_{abc} \bar c_c \frac{\delta }{\delta \bar c_b}
      + \epsilon_{abc} c_c \frac{\delta }{\delta  c_b}
 \nonumber \\&&
{
+\frac{i}{2}\tau_aL\frac{\delta \G}{\delta  L}
-\frac{i}{2}\bar L\tau_a\frac{\delta \G}{\delta \bar  L}
-\frac{i}{2}L^*\tau_a\frac{\delta \G}{\delta  L^*}
+\frac{i}{2}\tau_a\bar L^*\frac{\delta \G}{\delta \bar  L^*}
}
 \nonumber \\&& 
+ \epsilon_{abc} \Theta_{c\mu} \frac{\delta }{\delta \Theta_{b\mu} }
      + \epsilon_{abc} A^*_{c\mu} \frac{\delta }{\delta A^*_{b\mu}}
      + \epsilon_{abc} c^*_c \frac{\delta }{\delta  c^*_b} 
 - \frac{1}{2} \phi_0^* \frac{\delta }{\delta \phi^*_a}
\nonumber \\
&& +  
\frac{1}{2} \epsilon_{abc} \phi^*_c \frac{\delta }{\delta \phi^*_b} 
+ \frac{1}{2} \phi_a^* \frac{\delta }{\delta \phi_0^*} 
\Biggr)\G \, .
\label{bkgwi.1}
\end{eqnarray}
It is straightforward to prove that
\begin{eqnarray}
[{\cal S}_0,{\cal W}_0] = 0.
\label{fle.5}
\end{eqnarray}
%

\section{Subtraction Strategy}
\label{sec:sub}
The superficial degree of divergence in eqs. (\ref{pre.18})
or (\ref{pre.18p}) shows that the theory is not
renormalizable even if we invoke the hierarchy in
order to renormalize only the ancestor amplitudes.
This item has been considered at length by the
present authors. The extensive discussion is in
Ref. \cite{Bettinelli:2007zn}, where we argue
in favor of a particular subtraction procedure
which respects locality and unitarity at variance
with the algebraic renormalization which cannot
be implemented in the present case.
\par
To ferret out the procedure of the removal of 
divergences, eq. (\ref{bkgwi}) is used. Dimensional
regularization provides the most natural environment.
Let us denote by
\begin{eqnarray}
\G^{(n,k)}
\label{fle.6}
\end{eqnarray}
the vertex functional for 1-particle irreducible
amplitudes at $n$- order in loops where the
countertems enter with a total power $k$ in $\hbar$.
In dimensional regularization we can perform
a grading in $k$ of eq. (\ref{bkgwi}). Thus if
we have successfully performed the subtraction
procedure satisfying eq. (\ref{bkgwi}) up to
order $n-1$ the next order effective action
\begin{eqnarray}
\G^{(n)}=\sum_{k=0}^{n-1}\G^{(n,k)}
\label{fle.7}
\end{eqnarray}
violates eq. (\ref{bkgwi}) since the counterterm
$\hat \G^{(n)}$ is missing. The breaking term can be
determined by writing eq. (\ref{bkgwi}) at order 
$n$ at the grade $k\le n-1$ and then by summing over $k$.
One gets
\begin{eqnarray}
&&
{\cal W}_0\G^{(n)}
+ \frac{1}{2\Lambda^{(D-4)}}
\sum_{n'=1}^{n-1}\Bigl(
\frac{\delta\G^{(n-n')}}{\delta{K_0}}\Bigr)\Bigl(
\frac{\delta\G^{(n')}}{\delta{\phi_a}}\Bigr)
\nonumber\\&&
=\frac{1}{2\Lambda^{(D-4)}}
\sum_{n'=1}^{n-1}\Bigl(
\frac{\delta\G^{(n-n',n-n')}}{\delta{K_0}}\Bigr)\Bigl(
\frac{\delta\G^{(n',n')}}{\delta{\phi_a}}\Bigr).
\label{fle.8}
\end{eqnarray}
The first term in the LHS of eq. (\ref{fle.8}) has pole parts in $D-4$
while the second is finite, since the factors are of order $<n$, thus
already subtracted. The breaking term contains only counterterms
$\hat\G^{j}=\G^{(j,j)},~j<n$. This suggests the {Ansatz} that
the finite part of the Laurent expansion at $D=4$
\begin{eqnarray}
 \frac{1}{\Lambda^{(D-4)}}\G^{(n)}
\label{fle.9}
\end{eqnarray}
gives the correct prescription for the subtraction of the divergences;
i.e. one has to divide both members of eq. (\ref{fle.8}) by
$\Lambda^{(D-4)}$ and
remove only the pole parts (minimal subtraction). Thus the
counterterms have the form
\begin{eqnarray}
\hat\G^{(n)}= \Lambda^{(D-4)}\int \frac{d^Dx}{(2\pi)^D}{\cal M}^{(n)}(x)
\label{fle.10}
\end{eqnarray}
where the integrand is a local 
formal power series in
the fields, the external sources and their derivatives
(a local polynomial as far as the ancestor monomials
are concerned) and it possesses only pole parts
in its Laurent expansion at $D=4$.
\par
In conclusion the subtraction procedure relies on dimensional
regularization and it allows only one {extra} free parameter: $\Lambda$.
\par
In practice there are two ways to proceed in the regularization
procedure. One can use the forest formula and use minimal subtraction
for every (properly normalized) subgraph. It is possible, as alternative,
to evaluate the counterterms for the ancestor amplitudes and
then obtain from those all the necessary counterterms involving
the Goldstone boson fields $\vec\phi$.

\subsection{\huge $\gamma_5$}
\label{sec:g5}
Our subtraction procedure is based on minimal subtraction
on specifically normalized amplitudes. Thus the presence of 
$\gamma_5$ poses serious problems to the whole strategy. Our
approach is pragmatic. We introduce a new  $\gamma_{_D}$ 
which anticommutes with every $\gamma_\mu$ and at the same time
no statement is made about the analytic properties of
the trace involving $\gamma_{_D}$. Since the theory is not anomalous
such traces never meet poles in $D-4$ and therefore we
can evaluate the  traces at $D=4$. 
\par
%
\section{Physical observables}
\label{sec:obs}
The classical action 
of the nonlinearly realized electroweak model
(for gauge- and fermion fields)
has been modified in order to introduce 
mass term invariants. Moreover new fields ($c,\bar c,
c_0,\bar c_0,b, b_0$)
and sources have been added in order to perform
the quantization and  to establish the tools
necessary for the removal of divergences. 
The physical interpretation of the model has to
go through the standard selection of the physical
modes based on the Slavnov-Taylor linearized operator
${\cal S}_0$ in eq. (\ref{brst.14}). The unphysical
modes are the Faddeev-Popov ghosts, the scalar
components of the massive vector mesons
and of the photon and the Goldstone bosons. All these
modes have to conspire in order to give zero
contribution to the unitarity equation for physical states.
The STI is the essential tool
in order to guarantee that the theory has the right
unitarity property \cite{Ferrari:2004pd}. Beside the
fields, few sources and parameters have been introduced.
It is established in eq. (\ref{brst.14}) that these
auxiliary sources come in doublets 
\cite{Barnich:2000zw},\cite{Quadri:2002nh}
\begin{eqnarray}&&
{\cal S}_0V_\mu = \Theta_\mu
\nonumber\\&&
{\cal S}_0\phi_0^* = - K_0
\nonumber\\&&
{\cal S}_0\bar c = b.
\label{fle.11}
\end{eqnarray}
It can be shown that any physical relevant amplitude
can be separated into a part, where the doublets are
absent, plus terms  that are
irrelevant for the physical S-matrix elements.
These results are part of common knowledge for
the Lagrange multiplier $b$ \cite{PS}
and for the background
gauge field $V_\mu$ \cite{BFM}. It is a bit surprising that
the source $K_0$, coupled to the order parameter field
$\phi_0$, is an unphysical object. 
\par
The above argument shows also  that the constant $v$
introduced in eq. (\ref{pre.3}) is not a physical
parameter. In fact it is removed from the tree level
effective action  by rescaling $\vec \phi\to v\vec\phi$
and $K_0\to v^{-1}K_0$. The rescaling of  $\vec \phi$
is of no effect since it is a path integral integration
variable. The rescaling of $K_0$ has no effect on physical amplitudes
since it is a Slavnov-Taylor doublet.
\par
$\Lambda$ is a genuine new parameter that provides
the mass scale of the radiative corrections.
%
\section{Electric charge and gauge invariance}
\label{sec:charge}
Electric charge generates an exact symmetry, i.e. not
spontaneously broken. The associated local identity for the
generating functionals guarantees that longitudinal
polarizations of the electromagnetic potential decouple
from physical fields. It is worth to examine in detail
the corresponding identity for the vertex functional.

%
\begin{eqnarray}&& 
{\frac{1}{g'}}\Box b_0
+\Biggl(
- {\frac{1}{g'}}\partial^\mu  \frac{\delta }{\delta  B^\mu}
-{\frac{1}{g}}\partial_\mu \frac{\delta }{\delta A_{3 \mu}} 
-{\frac{1}{g}}\partial_\mu \frac{\delta }{\delta V_{3 \mu}}
\nonumber \\&&  
+ A_{2\mu} \frac{\delta }{\delta A_{1\mu}} 
-A_{1\mu} \frac{\delta }{\delta A_{2\mu}}
+iQL\frac{\delta }{\delta L}
-i\bar L Q\frac{\delta }{\delta \bar L}
+i Q R\frac{\delta }{\delta R}
-i\bar R Q\frac{\delta }{\delta \bar R}
\nonumber \\&&  
+ \phi_2 \frac{\delta }{\delta \phi_1}
-\phi_1\frac{\delta }{\delta \phi_2}
+  b_2 \frac{\delta }{\delta b_1}- b_1 \frac{\delta }{\delta b_2}
      +  c_2 \frac{\delta }{\delta  c_1}
      -  c_1 \frac{\delta }{\delta  c_2}
\nonumber \\&&  
      +  \bar c_2 \frac{\delta }{\delta \bar c_1}
      -  \bar c_1 \frac{\delta }{\delta \bar c_2} 
+  V_{2\mu} \frac{\delta }{\delta V_{1\mu}}
-  V_{1\mu} \frac{\delta }{\delta V_{2\mu}}
+ \Theta_{2\mu} \frac{\delta }{\delta \Theta_{1\mu} }
- \Theta_{1\mu} \frac{\delta }{\delta \Theta_{2\mu} }
\nonumber \\&&  
      + A^*_{2\mu} \frac{\delta }{\delta A^*_{1\mu}}
      - A^*_{1\mu} \frac{\delta }{\delta A^*_{2\mu}}
+ \phi_2^* \frac{\delta }{\delta \phi_1^*}
-\phi_1^* \frac{\delta }{\delta \phi_2^*}
      +  c^*_2 \frac{\delta }{\delta  c^*_1} 
      -  c^*_1 \frac{\delta }{\delta  c^*_2} 
\nonumber\\&&
-iQL^*\frac{\delta }{\delta L^*}
+i\bar L^*Q\frac{\delta }{\delta \bar L^*}
\Biggr)\G
= 0 \, ,
\label{charge.1}
\end{eqnarray}
where $Q$ is the electric charge of the component of the
multiplet.
Equation (\ref{charge.1}) is very important since it 
establishes gauge invariance (here, the decoupling of the
longitudinal polarizations of the photons from the physical
S-matrix elements). It is remarkable that the identity 
is a linear operator in a theory where physical unitarity
is guaranteed by BRST invariance, i.e. under nonlinear 
transformations; in fact the bilinear term $\G_{K_0}\G_{\phi_3}$
of eqs. (\ref{st.3}) and (\ref{bkgwi})
has cancelled out. Moreover it is surprising that
the equation takes such a simple form 
in the symmetric notation $(A_{3\mu},B_\mu)$. In fact, in terms
of the fields $Z_\mu,A_\mu$ in eq.(\ref{charge.2.mass})
the neutral boson part in eq. (\ref{charge.1}) takes the form
\begin{eqnarray}
- \frac{1}{g'}\partial^\mu  \frac{\delta }{\delta  B^\mu}
-\frac{1}{g}\partial_\mu \frac{\delta }{\delta A_{3 \mu}}
= - \frac{\sqrt{g^2+g^{'2}}}{gg'} \partial^\mu  \frac{\delta }{\delta  A^\mu}.
\label{charge.3}
\end{eqnarray}
The term $-\frac{1}{g}\partial_\mu \frac{\delta }{\delta V_{3 \mu}}$
takes into account that the field of the photon, as superposition
of $(A_{3\mu},B_\mu,\partial_\mu b_3)$, is modified by the
perturbative corrections. The latter are not present if only insertions of 
 ${\cal S}_0$-ivariant operators are considered, since $V_{a \mu}$, $\Theta_{a\mu}$ is 
 a BRST doublet, as discussed in Section \ref{sec:obs}.
%

\section{Conclusions}
\label{sec:concl}

In the framework provided by the nonlinear realization
of the gauge group $SU(2)\otimes U(1)$ for the electroweak
model new features show up, very interesting  both from the
phenomenological and  theoretical  point of view. The model
is nonrenormalizable; therefore the couplings with negative
dimension are not excluded and moreover the standard tools
for the subtraction of the divergences cannot be
applied.
\par
The discovery of a new LFE, which 
follows from the invariance of the path integral measure,
shows the way for a unique subtraction strategy of
the divergences without changing the number of tree-level
parameters apart from a common mass scale of the radiative
corrections. The algorithm is strictly connected with
dimensional regularization and symmetric subtraction  of
the pole parts in the Laurent expansion of the 1-particle
irreducible amplitudes.
\par
The same equation provides a hierarchy for the amplitudes:
those involving the Goldstone bosons can be derived from
those without. Thus also the superficial degree of divergence
of the graphs can be studied by means a new tool, the weak
power counting. The number of independent divergent amplitudes
turns out to be finite at any given loop order.  We argued that
this property is stable under the procedure of subtraction.
\par
The  weak power counting provides a way to solve the other problem
of limiting the anomalous couplings. In our opinion this seems
to be the only way to stabilize a nonrenormalizable model
under the process of making the amplitudes finite.
In the present case, $SU(2)\otimes U(1)$, two invariants appear
that contribute to the mass of the vector mesons. Thus the simple 
tree-level relation among masses and Weinberg angle is not working.
\par
The paper illustrates is a brief way all these points. We mention
the problem of $\gamma_5$ in dimensional regularization, where an
escape is permitted by the absence of any anomaly for the 
electroweak currents.
\par
We use the Landau gauge and  the fields in the symmetric basis.
This choice brings unexpected simplifications in the notations
and in the equations. Very interesting is the Ward identity associated
to the existence of the electric charge.

\end{document}